\documentclass[a4paper,11pt]{article}
\usepackage{pos}
\usepackage[export]{adjustbox}
\usepackage{mciteplus}
\title{Quantum correlation of neutral charmed mesons at BESIII}

\usepackage{ifthen}
\newboolean{articletitles}
\setboolean{articletitles}{true}

\author*[a,b]{Alex Gilman}

\onbehalf{for the BESIII collaboration}

\affiliation[a]{Department of Physics, University of Cincinnati,\\
  Cincinnati, Ohio 45221, USA}

\affiliation[b]{Department of Physics, University of Oxford,\\
Oxford OX1 3PU, UK\\
Presented at the 32nd International Symposium on Lepton Photon Interactions at High Energies, Madison, Wisconsin}

\emailAdd{alexander.leon.gilman@cern.ch}

\usepackage{xspace} 

\def\beq{\begin{equation}}
\def\eeq{\end{equation}}

\def\C      {\ensuremath{C}\xspace}

\def\psipp{\psi(3770)}

\def\DODO{\ensuremath{D^{0}}\ensuremath{\offsetoverline{D}^{0}}\xspace}
\def\DSTO{\ensuremath{D^{*0}}\xspace}

\def\aDSTO{\offsetoverline{D}^{*0}}

\def\DODO{\ensuremath{\Dz\Dzb}\xspace}
\def\DD{\ensuremath{\D\Db}\xspace}
\def\DSTDG{\ensuremath{\Dstar\Db\to {\gamma} \D\Db}\xspace}
\def\DSTDP{\ensuremath{\Dstar\Db\to{\pi^0} \D \Db}\xspace}
\def\DSTDSTEven{\ensuremath{\Dstar\Dstarb\to{\gamma\pi^0}\D \Db}\xspace}
\def\DSTDSTOdd{\ensuremath{\Dstar\Dstarb\to{\gamma\gamma/\pi^0\pi^0}\D \Db}\xspace}

\def\DSTD{\ensuremath{\Dstar\Db\xspace}}

\def\DSTDST{\ensuremath{\Dstar\Dstarb}\xspace}

\def\ee{e^+e^-}

\def\Ecm{E_{\text{cm}}}

\def\invfb{\text{ fb}^{-1}}

\def\GeV{\text{ GeV}}

\def\kaon    {{\ensuremath{\PK}}\xspace}

\def\K      {{\ensuremath{\kaon}}\xspace}
\def\deltakpi      {\ensuremath{\delta_{\K\pi}}\xspace}

\def\thebaroffset{0.1em}
\newcommand{\offsetoverline}[2][\thebaroffset]{\kern #1\overline{\kern -#1 #2}}%

\def\kaon    {{\ensuremath{\textit{K}}}\xspace}
\def\K      {{\ensuremath{\kaon}}\xspace}

\def\PD      {\ensuremath{D}\xspace}  
\def\Dbar    {{\ensuremath{\offsetoverline{\PD}}}\xspace}
\def\D       {{\ensuremath{D}}\xspace}
\def\Db      {{\ensuremath{\offsetoverline{{D}}}\xspace}}
\def\Dz      {{\ensuremath{\D^0}}\xspace}
\def\Dzb     {{\ensuremath{\Dbar{}^0}}\xspace}
\def\Dp      {{\ensuremath{\D^+}}\xspace}
\def\Dm      {{\ensuremath{\D^-}}\xspace}

\def\DpDm    {\ensuremath{\Dp {\kern -0.16em \Dm}}\xspace}
\def\Dstar   {{\ensuremath{\D^*}}\xspace}
\def\Dstarb  {{\ensuremath{\Dbar{}^*}}\xspace}
\def\Dstarz  {\DSTO\xspace}
\def\Dstarzb {\aDSTO\xspace}

\def\Ppsi        {\ensuremath{\psi}\xspace}  
\def\psipp  {{\ensuremath{\Ppsi(3770)}}\xspace}

\def\DstarzDstarzb    {\ensuremath{\Dstarz {\kern -0.07em \Dstarzb}}\xspace}

\def\deltaKpi      {\ensuremath{\delta_{K\pi}^\D}\xspace}

\def\C      {\ensuremath{C}\xspace}

\newcommand{\aunit}[1]{\ensuremath{\text{\,#1}}}       
    
\def\fb   {\ensuremath{\aunit{fb}}\xspace}
\def\invfb   {\ensuremath{\fb^{-1}}\xspace}
 
\abstract{BESIII has recently accumulated a large data sample near the \psipp production threshold corresponding to an integrated luminosity of 20\invfb. Neutral \DODO pairs produced at the \psipp are in a $C$-odd correlated state, providing a unique laboratory to measure the strong-phase differences between \Dz and \Dzb decays. These parameters are essential inputs to the study of CP violation in heavy-flavor physics, primarily in the determinations of the CKM angle gamma and charm-mixing parameters. These proceedings report new measurements of strong-phase differences in different neutral $D$ decays at BESIII, including new measurements of the strong phases in $\D\to K^+K^-\pi^+\pi^-$ decays. Additionally reported is the first observation of correlated $DD$ pairs produced at $\ee$ center-of-mass energies above the \psipp threshold, where $\ee\to\DSTD$ and $\ee\to\DSTDST$ processes also occur. These processes produce both $C$-odd and $C$-even correlated $\DD$ pairs, which allow for new measurement techniques to determine strong-phases from previously unused datasets.  }

\FullConference{%
  Lepton-Photon 2025\\
  Aug. 25- Aug. 29 2025\\
  Madison, Wisconsin, USA
}

\begin{document}
\maketitle


Hadrons produced in electron-positron collisions are constrained to respect the charge-conjugation eigenvalue of the photon, $C=-1$, as the strong and electromagnetic interactions mediate $\ee\to$ hadron processes. In the case of $\ee\to \DD$ production\footnote{The lack of charge superscripts here indicate the particles are produced as admixtures of $\Dz$ and $\Dzb$ mesons, with a similar convention taken for $\DSTO$ mesons.}, this constraint manifests through interference between the amplitudes of $\Dz$ and $\Dzb$ decays to the same final state\cite{Goldhaber:1976fp}. The observed interference is sensitive to the ratio of the amplitudes between the $\Dz$ and $\Dzb$ to the specific final state and the CP-conserving ``strong" phase between the amplitudes. These same parameters are important in the interpretation of CP violation in $B\to Dh$\cite{LHCbKShh} decays and the interpretation of neutral-charm meson mixing\cite{LHCbBinFlip}. Thus the measurements of strong phases from $\ee$ colliders running at open-charm thresholds provide crucial input to understanding the phenomena of CP violation and neutral-meson mixing.

\begin{figure}[h]
\centering
\includegraphics[width=0.45\textwidth]{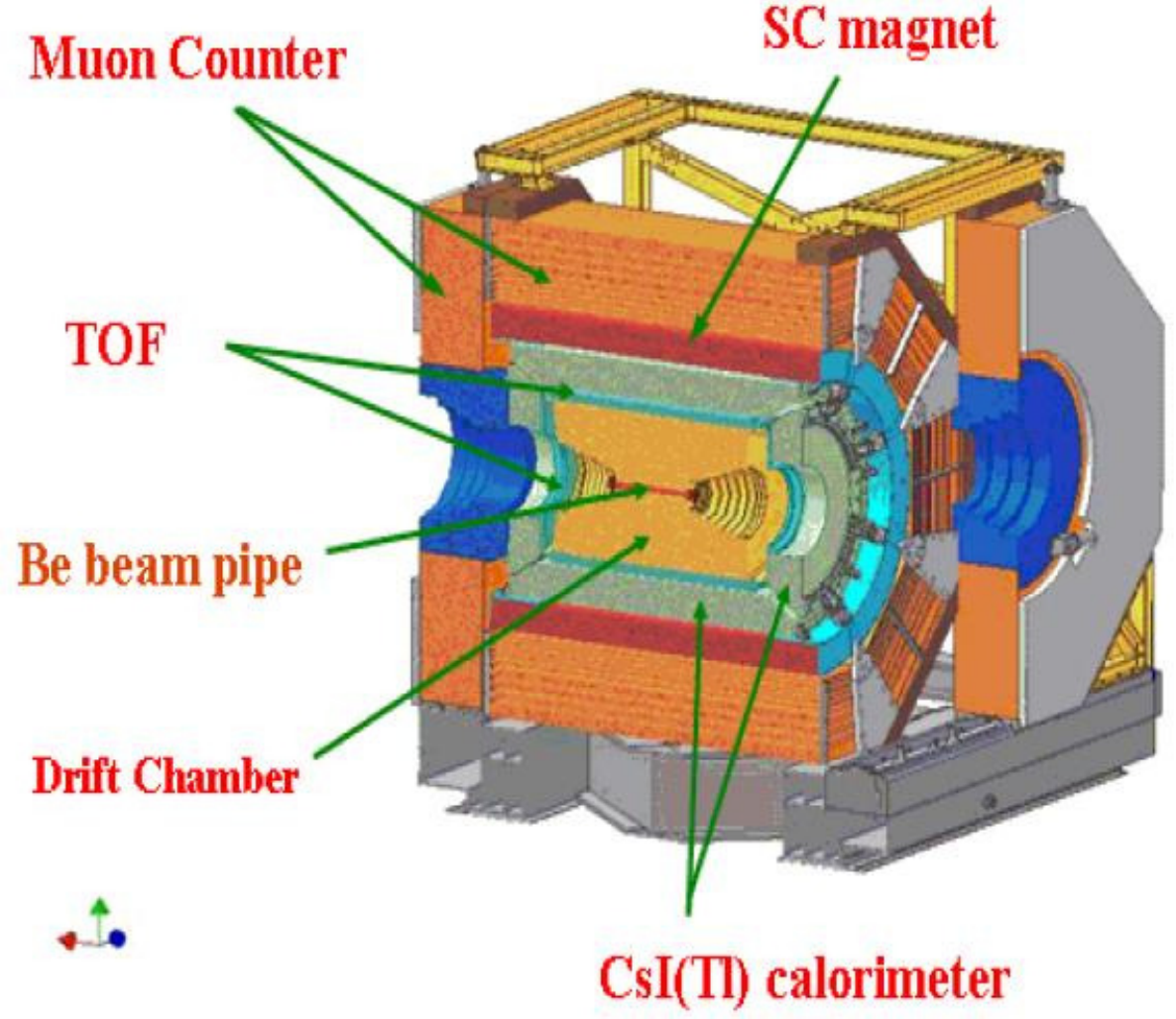}
\caption{Cross-sectional view of the BESIII detector.}
\label{fig:besiii}
\end{figure}

The BESIII experiment~\cite{BESIIIDetector} records data of symmetric $e^+e^-$ collisions provided by the Beijing Electron Positron Collider Mk. II \cite{BEPCII} (BEPCII). BEPCII produces collisions at center-of-mass energies of $2-5 \text{ GeV}$ with a design luminosity of $10^{33}$ $\text{cm}^{-2}\text{s}^{-1}$ achieved in April 2016. The BESIII detector covers $93\%$ of the full solid angle, and is equipped with gaseous tracking system, a plastic scintillator time-of-flight system for particle identification, a caesium-iodide calorimeter, and a resistive-plate-chamber muon system. A cross-sectional diagram of the BESIII Detector is shown in Fig.~\ref{fig:besiii}.
BEPCII and BESIII finished construction in 2008, upgrading on the original Beijing Electron-Positron Collider and BESII detector. BESIII began taking data in 2009, and has collected many large datasets in this energy regime\cite{BESIII:2020nme}. Most relevant to these proceedings is the $20.3\;\invfb$ sample collected at the \psipp threshold, whose collection was completed in 2024. This contains the previous largest sample at this energy, a sample of $2.9\invfb$ collected in 2011. Results reported in these proceedings also employ a sample corresponding to $7.3\invfb$ between center-of-mass energies of $4.13-4.23\GeV$ collected between 2013 to 2017.

At the $\psipp$ threshold, \DD pairs are constrained to a $CP=-1$ state, since the threshold production ensures that no other particles are produced\footnote{The only exception is $\ee\to\gamma\DD$ production, which is estimated to occur at a rate of $10^{-8}$ relative to the $\ee\to\DD$ production near the threshold, and so can be neglected.\cite{CLEOQC}}. In this case, the probability of the \DD pair decaying to a final state $X_1X_2$ relative to the independent probabilities $P(\Dz\to X_1)$ $P(\Dzb\to X_2)$ can be expressed as 
\begin{equation}\label{eq:codd_qc}
  	\frac{P(D^0\overline D^0\to X_1X_2)}{P(D^0\to X_1)P(\overline{D}^0\to X_2)}=1+ \left(r_D^{X_1}r_D^{X_2}\right)^2-2r_D^{X_1}r_D^{X_2}\cos\left(\delta_D^{X_1}+\delta_D^{X_2}\right),  
\end{equation}

where $r_D^X$ represents the ratio of amplitude magnitudes between $\Dz$ and $\Dzb$ decays to a final state $X$, and $\delta_D^X$ represents the strong phase between the two amplitudes. In order to determine the hadronic parameters of a decay $D\to X$, a variety of recoil $\Dz$ decay (or ``tag") modes must be reconstructed. These fall into three general categories:
\begin{enumerate}
    \item Flavor and quasi-flavor tags, which satisfy $r_D\approx0$
    \item CP and quasi-CP tags, which satisfy $r_D\approx1$ and $\delta_D\approx0,\pi$
    \item other indefinite tags, with a range of $r_D$ and $\delta_D$ values
\end{enumerate}

Examples of flavor tags include semileptonic $D^0$ decays, such as $D^0\to K^-e^+\nu_e$  or the quasi-flavor tag decay $D^0\to K^-\pi^+$. Quasi-CP eigenstate tags include $D^0\to K^+K^-$, $D^0\to K_S^0\pi^0$, and $D^0\to \pi^+\pi^-\pi^0$. 

For decays with three or more particles in the final state, the $r_D$ and $\delta_D$ parameters vary as a function of the final-state phase space. As measurements of these processes cannot resolve single points in phase-space, such multi-body decays are often analyzed with some coarse-graining of the phase-space. This requires assigning an effective amplitude ratio and strong-phase for the coarse-grained region of phase-space, as well as a coherence factor that accounts for the sensitivity loss due to the coarse-graining, often referred to with the symbol $\kappa$ or $R$. Alternative parameterizations of the phases, amplitudes, and coherence of the $D$ meson decays are also often employed in certain final states, such as $c_i$ and $s_i$ parameters, defined as

\begin{equation}\label{eq:cisi}
    c_i\propto \int_i d\Phi \;\left|\mathcal A_D\left(\Phi\right)\right|  \left|\mathcal A_{\overline D}\left(\Phi\right)\right|  \cos\left(\delta_D\left(\Phi\right)\right) \qquad      s_i\propto \int_i d\Phi \;\left|\mathcal A_D\left(\Phi\right)\right|  \left|\mathcal A_{\overline D}\left(\Phi\right)\right|  \sin\left(\delta_D\left(\Phi\right)\right),
\end{equation}

where the integral is performed of a subset of the total decay phase space $\Phi$ defined by $i$, and $A_{D}$ or $A_{\overline D}$ correspond to the $D^0$ or $\overline{D}^0$ decay amplitude to the final state of interest, respectively. Another commonly used alternative parametrization is the $CP$-even fraction $F_+$ of the decay phase-space of multi-body quasi-CP eigenstate decays.

In 2022-2024, BESIII recorded a data sample at $\Ecm=3.773 \GeV$ corresponding to an integrated luminosity of $17.4\invfb$, roughly five times larger than the previous largest sample of $2.9\invfb$ collected at the open-charm pair production threshold, which had been recorded by BESIII in 2011. The major increase in the number of quantum-correlated $\DD$ pairs provided by this sample will allow for significantly increased precision on strong-phase parameters in $D^0$ decays.

\begin{figure}[hbtp]
\centering{}
\begin{tabular}{ccc}
\includegraphics[width=0.30\linewidth,valign=c]{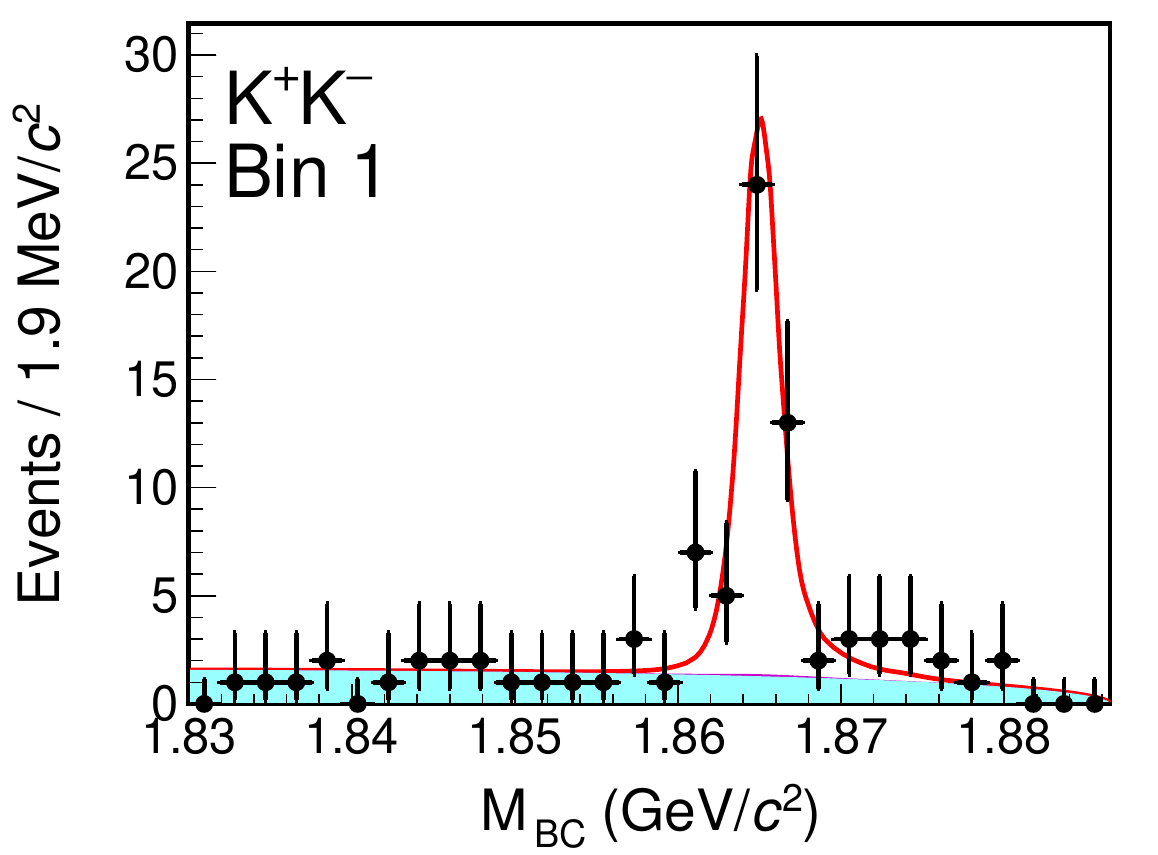}&\includegraphics[width=0.30\linewidth,valign=c]{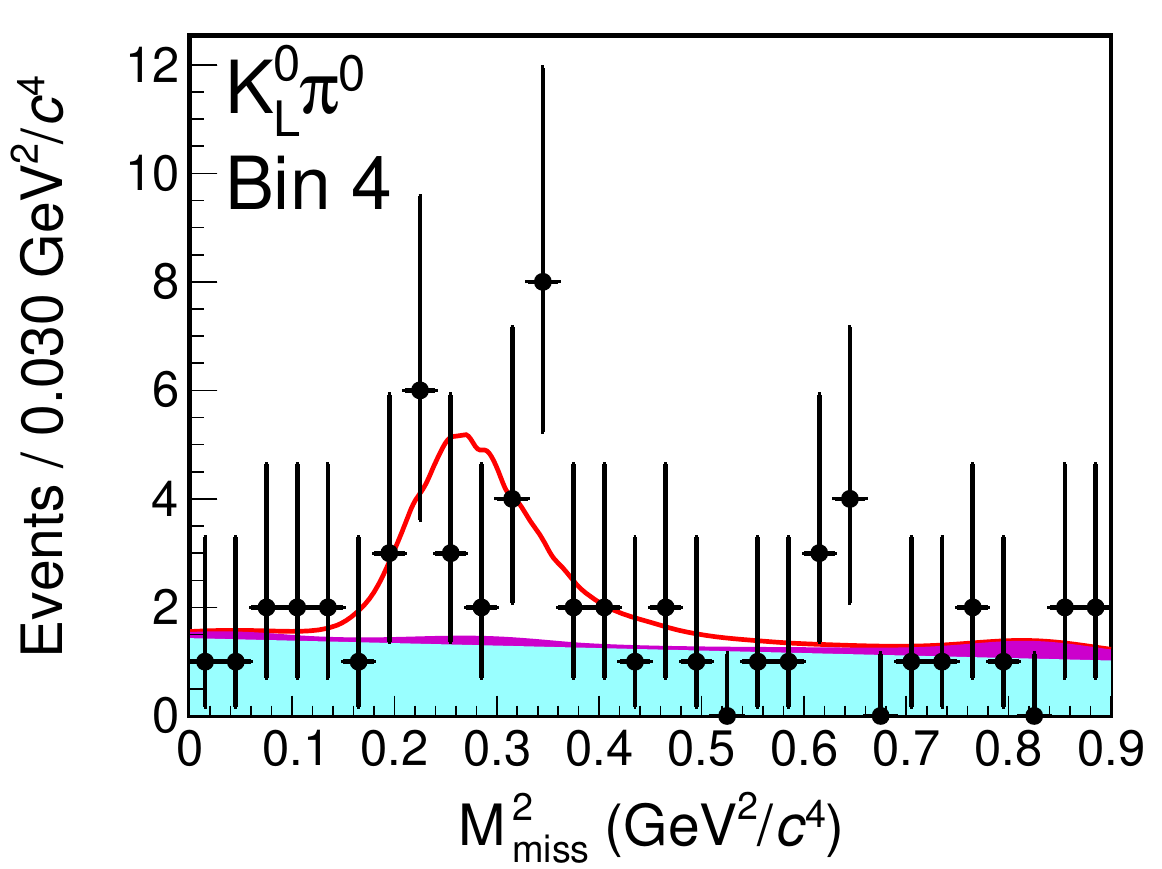} & \includegraphics[width=0.30\linewidth,valign=c]{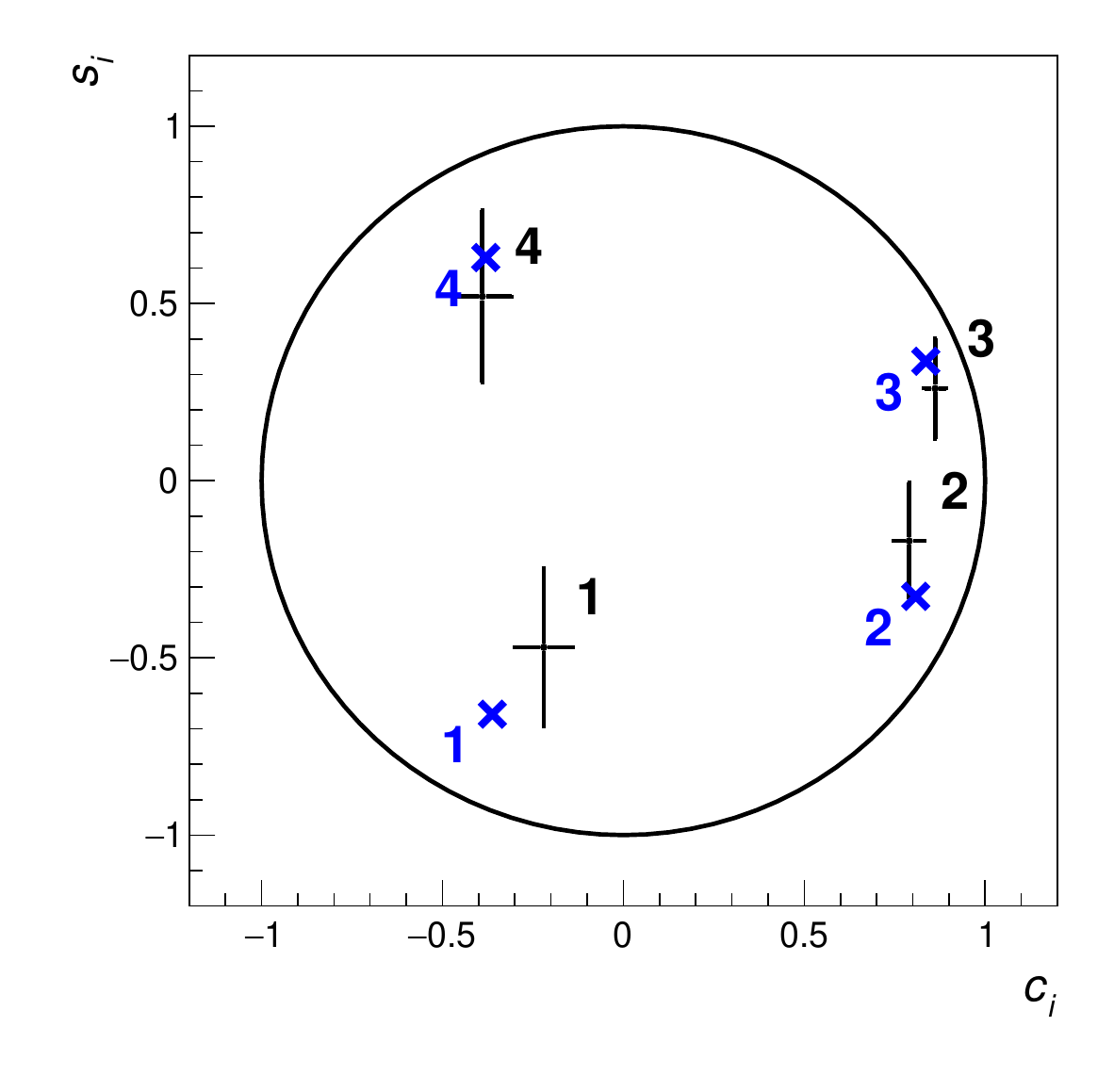}
\end{tabular}

\caption{(Left, Center) Examples of fits to determine the yields of $\DD\to K^+K^-\pi^+\pi^- \text{vs.}$ tag decay modes from Ref.~\cite{BESIII:2025wqu}. (Right) Determinations of the $c_i$ and $s_i$ parameters in Ref.~\cite{BESIII:2025wqu} as black points compared to predictions from an amplitude model\cite{LHCb:2023yjo} as blue crosses.  }
\label{fig:KKPiPiExample}
\end{figure}

The full data sample including the 2011 and 2022-2024 datasets has been employed in Ref.~\cite{BESIII:2025wqu} to produce the first measurement of $c_i$ and $s_i$ parameters in regions of the $D^0\to K^+K^-\pi^+\pi^-$ phase space designed to optimize sensitivity to the CKM angle $\gamma$ in decays of $B^\pm\to DK^\pm$ decays \cite{LHCb:2023yjo}, and an updated measurement of the $F_+$ of the $D^0\to K^+K^-\pi^+\pi^-$ decay integrated across the phase space. This analysis employs three quasi-flavour tag decay modes, ten CP and quasi-CP tag decay modes and the $K_{S/L}\pi^+\pi^-$ tag decay modes. Partially reconstructed $\D\to K^+K^-\pi^+\pi^-$ decays were also incorporated into the analysis, where one of the kaons was not successfully identified in the detector. Yields of $\DD\to K^+K^-\pi^+\pi^- \text{ vs.}$ the various tag decay modes are determined through fits to the beam-constrained mass of $\D\to K^+K^-\pi^+\pi^-$ ($M_{BC}$) for fully reconstructed events, and the missing mass squared ($M_\text{miss}^2$) for partially reconstructed events. Examples of these fits are shown in Fig.~\ref{fig:KKPiPiExample}. The yields are then analyzed simultaneously to determine the $c_i$ and $s_i$ parameters defined in Eq.~\ref{eq:cisi}, using reparameterised form of Eq.~\ref{eq:codd_qc}. The resulting determinations of these parameters are shown in Fig.~\ref{fig:KKPiPiExample}. The analysis also determines $F_+=\left(0.754\pm0.010\pm0.008\right)$ for the $D^0\to K^+K^-\pi^+\pi^-$ decay, and determines the branching fraction of $D^0\to K^+K^-\pi^+\pi^-$ to be $\left(2.863\pm0.028\pm0.045\right)\times10^{-3}$. These results have since been employed in a determination of the CKM angle $\gamma$ with $B^\pm\to Dh^\pm$ decays by the LHCb collaboration \cite{LHCb:2025fom}.

At energies above the open-charm pair production threshold, the processes $\ee\to X\DD $ may occur, where $X$ is an additional system of particles. If $X$ is also a $CP$-eigenstate, then it has been predicted that the $\DD$ system will exhibit quantum correlations \cite{Goldhaber:1976fp}. Additionally, if $X$ is $CP$-odd, the $\DD$ pair is expected to be constrained to a $C$-even state, with opposite behavior to $C$-odd constrained $\DD$ pairs, such as those produced at the threshold energy. However, no experimental verification of this has previously been undertaken.

\begin{figure}[hbtp]
\centering{}
\
\includegraphics[width=0.9\linewidth]{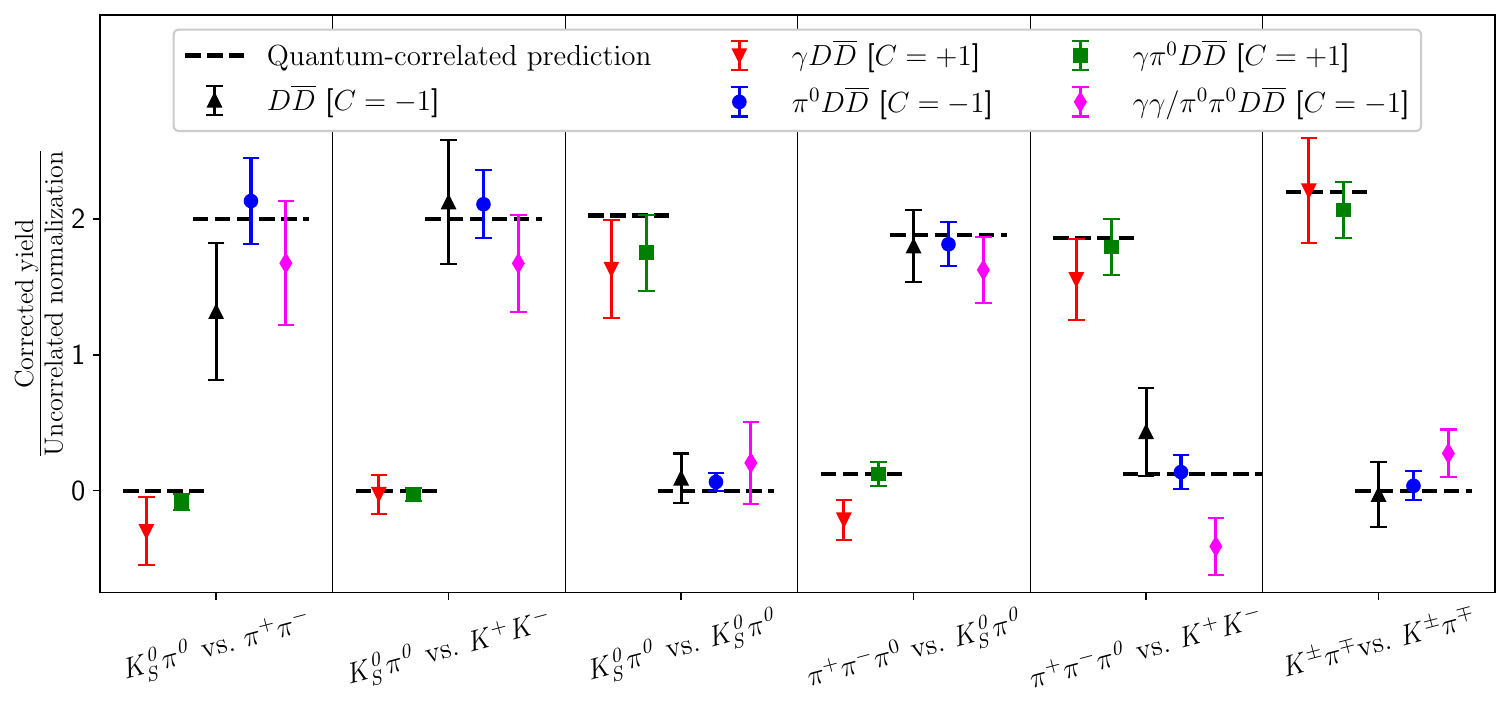}

\caption{Corrected yields of $\DD$ final states expected to be significantly affected by the presence of quantum correlations normalized to the expectation in the absence of correlations from Refs.~\cite{BESIII:2025pod,BESIII:2025xed}. The predicted ratios of yields in the presence of correlations are shown for comparison. }\label{fig:QCRatio}
\end{figure}

Using a data sample corresponding to an integrated luminosity of $7.3\invfb$ at center-of-mass energies ranging from $\Ecm=4.13-4.23\GeV$, the presence of correlations in $\DD$ pairs produced through $\ee\to X\DD $ processes has now been observed by the BESIII collaboration in the joint publications of Refs.~\cite{BESIII:2025pod,BESIII:2025xed}, specifically in $\DD$ pairs produced in five \mbox{$\ee\to X\DD$} production mechanisms expected to be \C-definite: \mbox{$\DD\;[\C=-1]$}, \mbox{$\DSTDG\;[\C=+1]$}, \mbox{$\DSTDP\;[\C=-1]$}, \mbox{$\DSTDSTEven\;[\C=+1]$}, and \mbox{$\DSTDSTOdd\;[\C=-1]$}. The yields are determined for $\DD$ final states that are expected to be forbidden or enhanced at a roughly twice the rate in the absence of correlations. Corrections are applied to the observed yields to account for reconstruction efficiencies and misidentification of the production mechanism hypothesis. The corrected yields for the final states expected to be significantly altered by the presence of correlations are compared to the corrected yields of the $\DD\to K^-\pi^+\text{ vs. } K^+\pi^-$ process, which is expected to be negligibly affected by the presence of the correlations. The ratios of corrected yields are shown in Fig.~\ref{fig:QCRatio}, which clearly demonstrates the presence of correlations in all of the observed final states.

The same data samples are employed to analyze decays of \mbox{$\DD\to K^\mp\pi^\pm\text{ vs. }Y$}, where $Y$ is one of the following $\left\{\pi^+\pi^-\pi^0, K^+K^-, \pi^+\pi^- ,K_S^0\pi^0, K_S^0\pi^+\pi^-\right\}$. Using a similar procedure to that employed to demonstrate the presence of quantum correlations in these samples, the effects of the correlations on the yields of these processes are studied to determine the strong phase between $D^0\to K^-\pi^+$ and $\overline D^{0}\to K^-\pi^+$ decays, $\deltakpi$. This results in a determination of \mbox{$\deltakpi=\left(192.8^{+11.0 + 1.9}_{-12.4 -2.4}\right)^\circ$}, where the first listed uncertainties are statistical and the second are systematic. This result is compared to the most precise measurement performed at the $\psi(3770)$ resonance, $\deltaKpi=\left(187.6^{+8.9 + 5.4}_{-9.7 -6.4}\right)^\circ$~\cite{deltaKpi}, where it is noted that the mixed $C$-even and $C$-odd sample allows for robust control of the systematic uncertainties. Given the similar sensitivities the two BESIII results are compared to determine $\deltaKpi=(189.2^{+6.9+3.4}_{-7.4-3.8})^\circ$.

The demonstration of $C$-even correlations also has implications for the study of neutral charm meson mixing. $C$-even correlated $\DD$ pairs are significantly more sensitive to the effects of mixing than $C$-odd correlated $\DD$ pairs\cite{Xing:1996pn,Bondar:2010qs,Rama:2015pmr,Naik:2021rnv}. In light of this, the sensitivity to charm-mixing parameters from hadronic final states at the proposed Super-Tau Charm facility\cite{STCF} is estimated in Ref.~\cite{BESIII:2025pod}, where it is demonstrated that sensitivity on par with the current world averages would be well within reach.

The results presented in these proceedings are a small subset of recent  BESIII results on quantum-correlated $\DD$ decays, with other results on the strong phases in $D^0\to \pi^+\pi^-\pi^+\pi^-$ \cite{BESIII:2024zco} using $2.9\invfb$ of data at the $\psi(3770)$ resonance, and  in $D^0\to \pi^+\pi^-\pi^0$\cite{BESIII:2024nnf}, $D^0\to K^+K^-\pi^0$\cite{BESIII:2024nnf}, and $D^0\to K_{S/L}^0\pi^+\pi^-$\cite{BESIII:2025nsp} decays using $7.9\invfb$ of data at the $\psi(3770)$ resonance, which both comprise a subset of the full $20.3 \invfb$ sample that has recently been collected. Many other results on strong phases are expected to follow employing the full $20.3 \invfb$ sample, which is projected to sufficiently improve the knowledge of strong phases in $D^0$ decays such that they do not limit the knowledge of the CKM angle $\gamma$ or charm-mixing parameters in the near future.

\bibliographystyle{LHCb}
\bibliography{main}

\end{document}